\documentclass[a4paper,11pt]{article}
\usepackage{pos}
\usepackage{xspace}
\title{Sibyll$^\bigstar$: ad-hoc modifications for an improved description of muon data in extensive air showers
}
 \ShortTitle{Sibyll$^\bigstar$}

\author*[a]{Felix Riehn}
\author[b,c]{Ralph Engel}
\author[d]{Anatoli Fedynitch}

\affiliation[a]{Instituto Galego de F\'isica de Altas Enerx\'ias (IGFAE),
Universidad de Santiago de Compostela, 15782 Santiago de Compostela, Spain}
\affiliation[b]{Karlsruhe Institute of Technology, Institute for Astroparticle Physics, 76021 Karlsruhe, Germany}
\affiliation[c]{Karlsruhe Institute of Technology, Institute of Experimental Particle Physics, 76021 Karlsruhe, Germany}
\affiliation[d]{Institute of Physics, Academia Sinica, Taipei City, 11529, Taiwan}

\emailAdd{friehn@lip.pt}
\emailAdd{ralph.engel@kit.edu}
\emailAdd{anatoli@gate.sinica.edu.tw}

\abstract{
Current simulations of air showers produced by ultra-high energy cosmic rays (UHECRs) do not satisfactorily describe recent experimental data, particularly when looking at the muonic shower component relative to the electromagnetic one. Discrepancies can be seen in both average values and on an individual shower-by-shower basis. It is thought that the muonic part of the air showers isn't accurately represented in simulations, despite various attempts to boost the number of muons within standard hadronic interaction physics. In this study, we investigate whether modifying the final state of events created with Sibyll~2.3d in air shower simulations can achieve a more consistent description of the muon content observed in experimental data. We create several scenarios where we separately increase the production of baryons, $\rho^0$, and strange particles to examine their impact on realistic air shower simulations.
Our results suggest that these ad-hoc modifications can improve the simulations, providing a closer match to the observed muon content in air showers.
One side-effect of the increased muon production in the considered model versions is a smaller difference in the predicted total muon numbers for proton and iron showers.
However, more research is needed to find out whether any of these adjustments offers a realistic solution to the mismatches seen in data, and to identify the precise physical process causing these changes in the model.
We hope that these modified model versions will also help to develop improved machine-learning analyses of air shower data and to estimate sys.{} uncertainties related to shortcomings of hadronic interaction models.

}

\ConferenceLogo{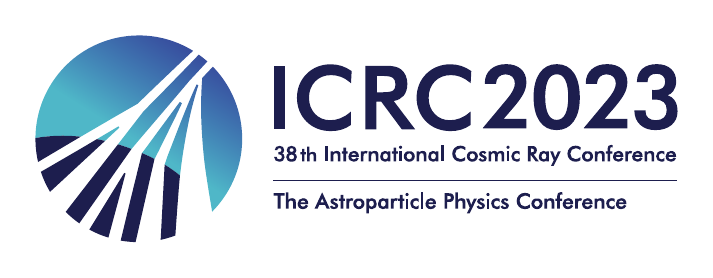}

\FullConference{%
38th International Cosmic Ray Conference (ICRC2023)\\
  26 July - 3 August, 2023\\
  Nagoya, Japan}

\def\xmax{\ensuremath{ X_{\rm max}}\xspace}
\def\mxmax{\ensuremath{\langle X_{\rm max} \rangle}\xspace}

\def\nmu{\ensuremath{N_\mu}\xspace}

\def\mlnmu{\ensuremath{\langle \ln N_\mu \rangle}\xspace}

\def\mnmu{\ensuremath{\langle N_\mu \rangle}\xspace}

\begin{document}
\maketitle

\section{Introduction}

\noindent
This study addresses the 'Muon Puzzle'~\cite{Albrecht:2021cxw,WHISPicrc2023}, a key challenge in the interpretation of extensive air shower (EAS) data. Specifically, we focus on the discrepancy between observed and predicted muon numbers in EAS, along with the associated inconsistency between the depth of shower maximum (\mxmax) and the ground-level signal~\cite{AugerXmaxDNN,Vicha:2022zvv}. In pursuit of gaining better understanding of this problem, we explore ad-hoc modifications of the hadronic interaction model Sibyll~2.3d~\cite{Ahn:2009wx,sib23eas,sib23flux}. While numerous standard~\cite{Ostapchenko:2013pia,Drescher:2007hc,Pierog:2006qv,Grieder:1973x1,Baur:2019cpv} and exotic~\cite{Farrar:2013sfa,AlvarezMuniz:2012dd,Rybczynski:2019exi,Anchordoqui:2016oxy,Anchordoqui:2022fpn} mechanisms have been proposed to address this problem, our study specifically focuses on quantitatively evaluating several conventional mechanisms. For this purpose, we employ a customized version of Sibyll, designed for direct use in realistic EAS simulations. All modifications of the model satisfy the constraint that, on an event-by-event level, all fundamental constraints such as energy-momentum and quantum numbers of relevance are conserved. Furthermore, existing experimental data from particle physics experiments are considered as guiding input, limiting some of the changes to energy and phase space regions for which no such data exist.

\section{Sibyll$^\bigstar$}

\noindent
We construct the custom models by modifying events generated with Sibyll~2.3d. Once the initial event generation is complete, we start by letting all hadronic resonances with a shorter lifetime than that of K$^0_{\rm s}$ decay, except for $\pi^0$. We then go through the event's particle list to identify appropriate pairs or triples of pions, only considering the five nearest neighbors for every pion. If the combined invariant mass of the selected pions is sufficient and the sampling criterion is fulfilled, we exchange them with a pair of new particles that together have the same total momentum, invariant mass, and charge. We calculate the final momenta from the invariant mass, the new particles' masses, and a minor transverse momentum sourced from an exponential distribution in transverse mass. In these altered events, we maintain conservation of energy, momentum, and charge, although (iso)spin conservation is not upheld.

The acceptance rate of these particle exchanges hinges on the total center-of-mass (CM) energy ($\sqrt{s}$) and the fraction of longitudinal momentum $x_{\rm F}$ (defined as $p_{z}/p_{z,\mathrm{max}}$, with momenta $p$ in the CM frame) of the initial particle. The probability of exchanging particles is parameterized by
\begin{equation}
  P_i ~=~ P_{i,0} \cdot |x_{\rm F}|^{\epsilon_i} \cdot f(\sqrt{s}, E_{\rm thr}) \ .
\end{equation}
The emphasis given to forward or central particles depends on the chosen value for the exponent $\epsilon_i$ in the $x_{\rm F}$-dependence. If $\epsilon_i=0$, all particles receive equal weight, preserving the original distribution's shape in longitudinal phase space. As $\epsilon_i$ approaches 1, the forward part of the $x_{\rm F}$-spectrum is enhanced.

The energy dependence of $f(\sqrt{s},E_{\rm thr})$ is logarithmic. It is set such that the rate is precisely zero below a threshold energy $E_{\rm thr}$ and reaches the nominal $P_{0,i}$ at lab energies of $10^{19}\,$eV ($1.37\times10^5\,$GeV in CM frame). The threshold energy $E_{\rm thr}$ is set at $5\,$GeV.

This parameterization of energy dependence represents a very gradual change in particle production, from no change at low energies, where fixed target experiments effectively limit the entire phase space, to LHC energies, where only the central region is well constrained, up to the UHECR energy scale (with no lab experiment constraints). Overall, this spans five orders of magnitude in energy, and the modification scales from zero to one. As an alternative, we allow a more drastic increase of the exchange rate towards high energies (above $13\,$TeV in CM frame) represented as $ P_{i} \to P_{i,0} + P_{i,\mathrm{HE}} \cdot f_{\rm HE}(\sqrt{s},13\,\mathrm{TeV})$. This mode depicts a swift transition to new physics beyond the LHC scale. Lastly, we apply this algorithm for all projectiles or only for mesons.


\begin{figure}[h]
  \centering
  \includegraphics[width=\textwidth]{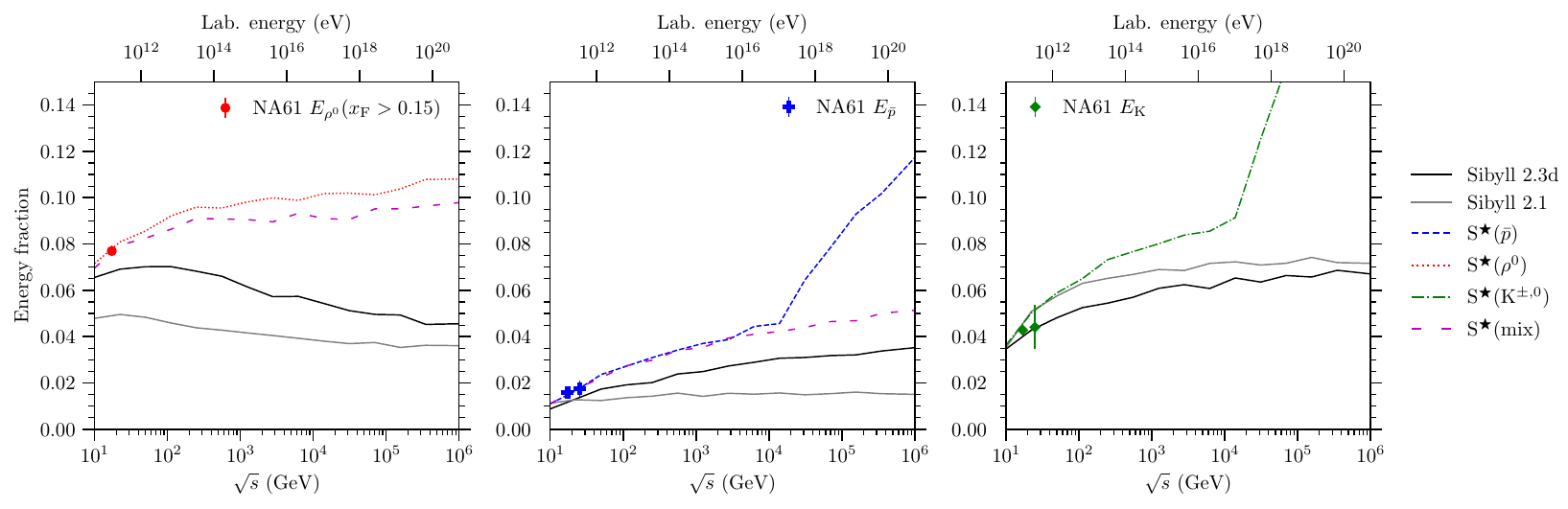}
  \caption{Fraction of projectile energy carried by $\rho^0$, anti protons and charged kaons in $\pi^-$C collisions~\cite{Aduszkiewicz:2017anm,NA61SHINE:2022tiz}. Lines are Sibyll~2.3d, Sibyll~2.1 and different variants of Sibyll$^\bigstar$.}
  \label{fig:en-na61}
\end{figure}

\begin{figure}[h]
  \centering
  \includegraphics[width=\textwidth]{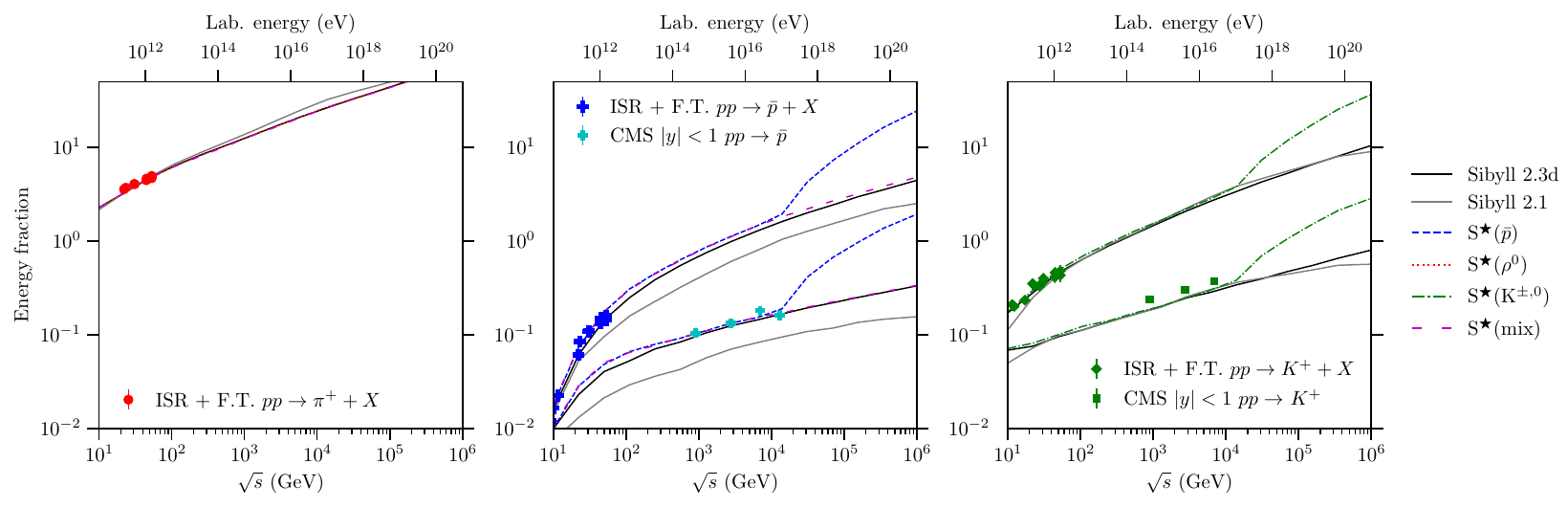}
  \caption{Multiplicities of $\pi^+$, anti protons and charged kaons in pp collisions~\cite{Sirunyan:2017zmn,Chatrchyan:2012qb,Albini:1975iu}. Lines are Sibyll~2.3d, Sibyll~2.1 and different variants of Sibyll$^\bigstar$.}
  \label{fig:mult}
\end{figure}

Using the previously discussed algorithm, we construct different Sibyll~2.3d variants with an aim to enhance muon production in EAS. We select three distinct modifications known for their efficacy in muon production: $\rho^0$ production, baryon anti-baryon pair-production, and kaon production enhancement~\cite{Ostapchenko:2013pia,Drescher:2007hc,Pierog:2006qv,Grieder:1973x1,Anchordoqui:2022fpn}. These variants are denoted as S$^\bigstar$($\rho^0$), S$^\bigstar$($\bar{p}$), and S$^\bigstar$(K$^{\pm,0}$).

In the $\rho^0$ variant, $\pi^0$ are directly substituted with $\rho^0$. For the baryon pair and kaon pair variant, charge-neutral combinations of two or three pions are replaced with $p\bar{p}$ or $n\bar{n}$ pairs, and K$^+$K$^-$ or K$^0$ $\bar{\mathrm{K}}^0$ pairs respectively. We adjust the parameters for each variant to align the energy in the corresponding component with NA61 measurements~\cite{NA61SHINE:2022tiz,Aduszkiewicz:2017anm} (refer to Fig.\ref{fig:en-na61} and Fig.~\ref{fig:mult}).

Producing $\rho^0$ significantly impacts muon production in EAS because $\rho$ mesons can form directly from the pion projectile. This mechanism doesn't hold for proton projectiles, hence for the $\rho^0$ variant, only meson projectile interactions are modified. However, the modifications for baryon pair and kaon production apply to any projectile and include a rapid increase at energies beyond the LHC. The parameters are adjusted so that the energy fraction carried by all hadrons except neutral pions at $10\,$EeV is approximately the same across all variants ($\approx 0.82$).

Furthermore, we create a fourth variant using both $\rho^0$ and baryon pair production (S$^\bigstar$(mix)). In this scenario, we choose a more moderate increase of $\rho$ production and dismiss the rapid increase of baryon production at high energies. The parameters of the different variants are detailed in Tab.~\ref{tab:variants}.

\begin{table}[h]
  \caption{Parameters in different variants of Sibyll$^\bigstar$. \label{tab:variants}}
  \begin{center}
    \renewcommand{\arraystretch}{1.1}
    \begin{tabular}{ccccc}
      \hline
      Label & $P_{i,0}$ & forward weight $\epsilon_i$ & projectiles & $P_{i,\mathrm{HE}}$ \\     
      \hline
      $\rho^0$              & 0.8 & 0.3 & mesons & -  \\
      $\bar{p}$           & 0.5 & 0.7 & all & 0.25  \\
      $K^{\pm,0}$           & 0.5 & 0.8 & all & 0.3  \\
      $\rho$-mix          & 0.8 & 0.4 & mesons & -  \\
      $\bar{p}$-mix       & 0.5 & 0.7 & all & -  \\
      \hline
    \end{tabular}
  \end{center}
\end{table}

Note that the mechanism driving the effectiveness of both strangeness enhancement and increased baryon pair production in raising the muon number is fundamentally the same: quantum number conservation (strangeness and baryon number, respectively). The key difference is that baryon production works effectively at all energies (no nucleon decay), while strangeness is only conserved in EAS at high energies, where kaon decay is negligible. Note also that since hyperons do not decay into kaons, enhanced hyperon production is part of the variant with enhanced baryon production.

\section{EAS predictions}

\begin{figure}[h]
  \centering
  \includegraphics[width=0.48\textwidth]{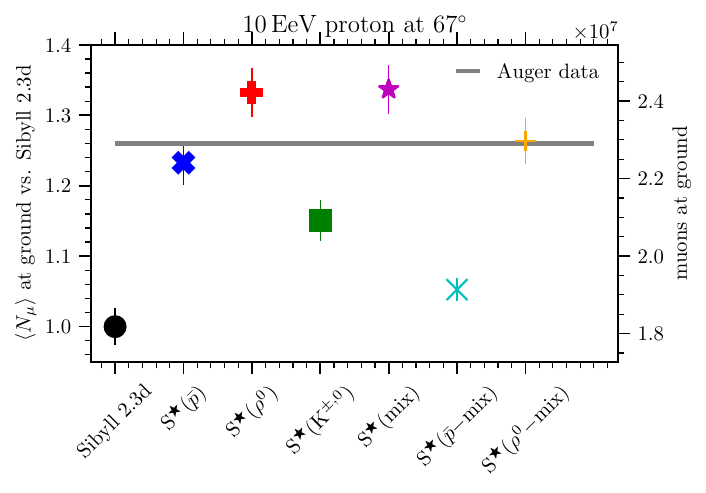}
  \hfill
  \includegraphics[width=0.48\textwidth]{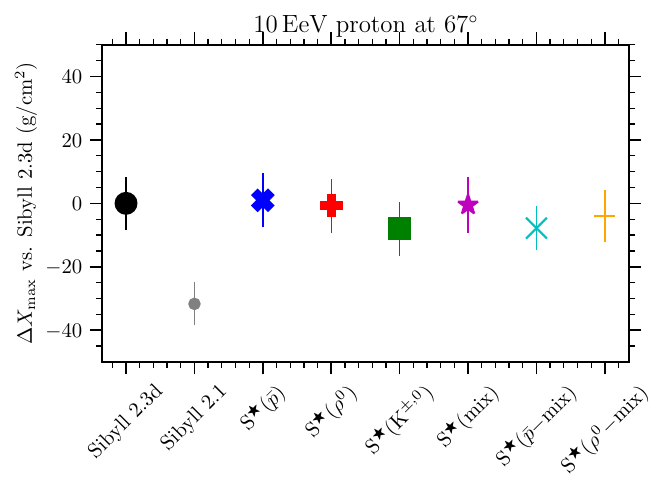}
  \caption{\nmu and \xmax for proton showers at $67^\circ$ across Sibyll variants. The left panel shows a substantial increase, up to 35\%, in the number of muons for the mixed and $\rho$ variant. However, the variation on the shower maximum between Sibyll~2.3d and its variants is less than $7\,$g$/$cm$^2$. The grey line represents the required increase in muon count to align with the data from the Pierre~Auger~Observatory.}
  \label{fig:nmu-xmax-change}
\end{figure}

\noindent
In our EAS predictions, we carried out full air shower simulations using CORSIKA\footnote{CORSIKA~v7.7420~\cite{Heck98a}; The magnetic field and observation level are set to the values of the site of the Pierre~Auger~Observatory. Low-energy hadronic interactions were modeled using FLUKA~2021.2.9~\cite{fluka2014}.} for each S$^\bigstar$ variant. We compared the resulting average depth of shower maximum (\mxmax) and the average number of muons at ground level (\mnmu) with simulations using the unmodified Sibyll~2.3d. These simulations were set to mirror conditions at the Pierre~Auger~Observatory site. The results for a primary proton with energy $10\,$EeV and an incident zenith angle of $67^\circ$ are presented in Fig.~\ref{fig:nmu-xmax-change}. These results show a notable increase in the number of muons, while \mxmax remains largely unaffected.

\begin{figure}[h]
  \centering    
  \includegraphics[width=0.6\textwidth]{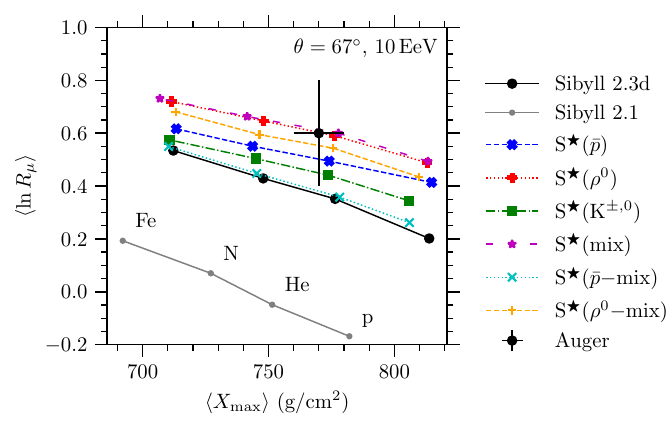}
  \caption{Comparison of the predicted values for \mxmax and \mnmu from various Sibyll$^\bigstar$ variants to the measurements obtained from the Pierre~Auger~Observatory.~\cite{PierreAuger:2021qsd}}
  \label{fig:rmu-vs-xmax}
\end{figure}

In the analysis shown in Fig.~\ref{fig:rmu-vs-xmax}, we compare the predicted values of \mxmax and \mlnmu from proton, helium, nitrogen, and iron primaries with the measurement from the Pierre~Auger~Observatory~\cite{PierreAuger:2021qsd}. Our findings indicate that only the $\rho^0$ and the mixed $\rho^0$ baryon-pair variant yield a sufficient number of muons to match the levels shown in the data. 

The model predictions in Fig.~\ref{fig:rmu-vs-xmax} for p, He, N and Fe all fall on a line. The reason is that per the superposition model, \mxmax and \mlnmu, have the same dependence on the primary mass (linear in $\ln A$), e.g.\ $\mlnmu(A,E)~=~(1-\beta)\ln A \, + \, \mlnmu(1,E)$, where $\beta$ is the exponent in the energy dependence of the number of muons for proton primaries, that is $\mlnmu(1,E)~=~\beta \ln E$. In simplified models à la Heitler-Matthews we have
\begin{equation}
    N_\mu = A\cdot  \left(\frac{E}{A\cdot E_{\rm dec}}\right)^\beta
\end{equation}
with $\beta$ being computed as $\ln N_{\rm ch}/\ln N_{\rm tot}$, where $N_{\rm ch}$ and $N_{\rm tot}$ denote the multiplicities of charged and all pions, respectively~\cite{Matthews:2005sd}. More broadly, $\beta$ is associated with the fraction of hadrons that carry enough kinetic energy to re-interact rather than decay. The slopes in Fig.~\ref{fig:rmu-vs-xmax} decreases from Sibyll~2.1 towards the variants with the highest levels of muon production. This behavior is common to all variants S$^\bigstar$. The greater a fraction of energy is kept in hadrons, the larger the increase in the number of muons and the larger $\beta$, and consequently, the less pronounced is the separation of the primary masses in \mnmu.

\begin{figure}[h]
  \centering
  \includegraphics[width=\textwidth]{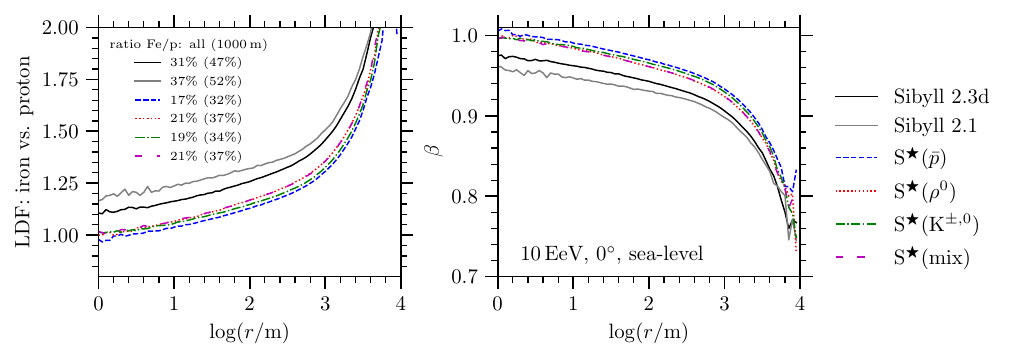}
  \caption{
    Ratio between the average number of muons in iron and proton showers (left panel) and the slope $\beta$ (right panel). Both are shown as a function of the distance from the shower axis. The inset numbers in the left panel show the ratio of the total numbers of muons between iron and proton showers and the ratio of the muon densities at $1000\,$m in parentheses.
  }
  \label{fig:beta-vs-r}
\end{figure}

The data clearly favor a larger muon content at high energy leading to a reduction in the mass resolution. However Fig.~\ref{fig:beta-vs-r} shows that for specific experiments the situation may be not as dramatic, as the separation between primary masses (respectively $\beta$) varies with the distance from the shower axis. The reason is that muons at different lateral distances are dominated by different phases of the shower development~\cite{sib23eas,Cazon:2022msf,MULLER2018174,Cazon:2012ti,Maris:2009uc}.

\section{Inclusive fluxes}

\noindent
We employed the MCEq code~\cite{Fedynitch:2015zma} for a comparative analysis of the atmospheric muon and neutrino fluxes as predicted by the S$^\bigstar$ versions against the original Sibyll~2.3d. Considering the modification (S$^\bigstar(\rho^0)$) impacts secondary pion interactions and (S$^\bigstar(\bar{p})$) affects the production of secondary baryons, both of which influence the shower development in deeper atmospheric layers, we didn't foresee changes to inclusive fluxes, which primarily depend on the air shower's early stages. Our findings confirm no significant effects. However, since $K^{\pm}$ decays into muons and neutrinos, for the S$^\bigstar$(K${^\pm,0}$) model, we identified a minor increase in muon fluxes of around 5\% at tens of TeV and PeV energies, and up to a 20\% increase in atmospheric neutrino fluxes.

\section{Discussion}

\noindent
In this study, we explored custom variants of the Sibyll~2.3d model, aiming to boost muon production in extensive air showers. Our focus centered on three modifications: $\rho^0$ production, baryon anti-baryon pair-production, and kaon production enhancement. Significantly, the $\rho^0$ and mixed $\rho^0$ baryon-pair enhancements effectively aligned with the observed muon production data from the Pierre~Auger~Observatory at $10\,$EeV. In contrast, elevating strangeness or baryon production proved insufficient, even when amplified to extreme levels. However, our results don’t definitively eliminate these scenarios, especially considering that our implementation does not permit the production of leading strangeness (e.g. $p \to \Lambda^0 \to \mathrm{K}^+ \, +\, $n).

Our simulations show that these modifications increase the number of muons, while the average depth of shower maximum remains largely unaffected. Another observation is that the degree of primary mass separation through muon measurement is predicted to be smaller in the modified models than in the original Sibyll~2.3d and other interaction models as along as the total number of muons is concerned. This is also expected within the Heitler-Matthews model~\cite{Matthews:2005sd} as more energy is kept in the hadronic channel in each interaction, increasing the exponent $\beta$. On the other hand, the muon density at sufficient distance from the shower core (for example, 1000\,m) is a very well suited observable for composition studies. 

Moreover, despite certain modifications leading to increased muon fluxes, inclusive fluxes - primarily dependent on early air shower stages - remained largely unchanged. Only the S$^\bigstar$(K${^\pm,0}$) model showed a minor increase in muon fluxes and a noticeable increase in atmospheric neutrino fluxes, due to the decay of $K^\pm$ into muons and neutrinos. 

The provided variants of Sibyll~2.3d can be used to train machine-learning models like deep neural networks to have a better description of the measurements and to estimate systematic uncertainties stemming from shortcomings of modeling hadronic multiparticle production.

{\small
\paragraph{Acknowledgements} The authors acknowledge many fruitful discussions with colleagues of the Pierre~Auger and IceCube Collaborations. They have used computing resources provided by the Academia Sinica Grid Computing Center (ASGC), supported by the Institute of Physics at Academia Sinica. FR and RE are supported in part by the European Union’s Horizon~2020 research and innovation programme under the Marie~Skłodowska-Curie grant agreement No.{} 101065027 and by BMBF grant No.{} 05A2023VK4, respectively.

\let\OLDthebibliography\thebibliography
\renewcommand\thebibliography[1]{
  \OLDthebibliography{#1}
  \setlength{\parskip}{0pt}
  \setlength{\itemsep}{0pt plus 0.3ex}
}


\providecommand{\href}[2]{#2}\begingroup\raggedright\endgroup

}

\end{document}